# A yellow giant, a peculiar A-type dwarf and an interstellar dust cloud unravel the eclipsing binary TYC 4481-358-1

Norbert Hauck

Bundesdeutsche Arbeitsgemeinschaft für Veränderliche Sterne e.V. (BAV), Munsterdamm 90, 12169 Berlin, Germany : hnhauck@yahoo.com

*A first solution for the eclipsing binary TYC 4481-358-1 has been found by combining the results of BVIc-photometry with known stellar models and stellar spectral energy distributions (SEDs). The binary shows total and annular eclipses in a circular 90-days orbit. Masses, radii and effective temperatures have been derived: about 3.01 Msun, 14.29 Rsun and 4950 K for the giant, and about 2.39 Msun, 2.39 Rsun and 9600 K for the dwarf. The peculiar dwarf is discolored and shows a Teff - passband dependence from 9000 K in the Ic to 9800 K in the B band. The giant is in its final stage of core helium burning. The interstellar extinction in our line of sight is considerable (Av about 1.44 mag at an elevated Rv of 3.58).*

The eclipsing binary nature of TYC 4481-358-1 (GSC 04481-00358) has been discovered by S. Otero and is described in the VSX database of the AAVSO [1]. Total eclipses in a 90-days orbit have been reported there, based on photometric data of the ASAS-SN and NSVS sky surveys. From the GAIA mission's third data release (DR3) we know the distance to the binary of 617 (612 - 623) pc as well as a preliminary value of the total full amplitude of the radial velocity of 69.71 km/s. No spectral type information has been found in the literature.

New photometric data in the passbands BVIc have been collected with a remotely controlled 17-inch CDK reflector telescope in New Mexico, USA. The *Binary Maker 3* (BM3) program has been used for the modeling procedure. Fig. 1 contains 62 of our binned data points and the light curve computed for an effective wavelength $\lambda_{eff}$ of 440 nm ($\sigma_{FIT}$ = 3.2 mmag). The circular orbit generates total eclipses in the primary, and annular eclipses in the secondary minimum. Additionally, for 36 data points of the primary minimum taken from the photometry of the TESS mission's Full Frame Images (Huang et al. [2]) the light curve has been computed at a $\lambda_{eff}$ of 790 nm ($\sigma_{FIT}$ = 0.28 mmag) for orbit synchronized, i.e. here rather slowly rotating stars (see Fig. 2).

The effective surface temperature $T_{eff}$ of the giant component of the binary has been calculated from infrared photometry data of the 2MASS mission in the passbands J, H and K (resp. $K_S$), after subtraction of the dwarf's flux contribution. The extinctions (A) in J, H and K have been converted to the extinction in passband V ($A_V$) by using the coefficients $A_{J,H,K}/A_V$ of the extinction law of Wang & Chen [3]. Thereby the ratio of two extinction-free color indexes, e.g. $(J-H)_0 /(J-K)_0$, can be calculated independently of $A_V$. In the corrected spectral energy distributions (SED) of the standard stellar library II of T. Lejeune et al. [4] these ratios fit to a $T_{eff}$ of 4920 (+140/–80) K for the giant component (at log g 2.6 and [Fe/H] 0.0).

The interstellar extinction parameter $R_V$ of about 3.58 has been calculated with the empirical equation of Fitzpatrick & Massa [5] : $R_V$ = –1.36 ($E_{K-V}/E_{B-V}$) – 0.79, by using the extinction-free color indexes of the SED of Lejeune et al. [4] for getting the color excesses E. Since total to selective extinction ratio $R_V = A_V/E_{B-V}$, we now also get the extinction $A_V$ of about 1.44, which is clearly above the normal values for the known distance of the binary. Equally elevated is our (dust properties dependent) $R_V$, compared to an $R_V$ of 3.16 ± 0.15 for the normal galactic diffuse interstellar medium (ISM) (see Wang & Chen [3]). Hence both $A_V$ and $R_V$ indicate the presence of an interstellar or circumbinary dust cloud in our line of sight.

At the known distance, extinction and a $T_{eff}$ of 4950 K our giant star gets a calculated size of 14.29 $R_\odot$, which nicely fits into the stellar model of Ekström et.al. [6] at solar metallicity (Z = 0.014) without rotation for a mass of 3.01 $M_\odot$ at the final stage of core helium burning. The light curve fit in the BM3 then allows the determination of the orbital radius of 148.7 $R_\odot$ and the total binary mass of 5.40 $M_\odot$. Simultaneously, our dwarf component of 2.39 $R_\odot$ and 2.39 $M_\odot$ fits into the main sequence of this stellar model at a $T_{eff}$ of 9600 K. And, as it should be, both stars are coeval, i.e. they have an age of 416 Myr in the stellar model. The calculated peak-to-peak amplitude of the giant's radial velocity of 73.55 km/s is in line with the preliminary and hence adopted lower limit value of 69.71 km/s from GAIA (DR3).

Our photometric BVIc data fits have been compared with those obtained from the Full Frame Image (FFI) precision photometry of the TESS mission (see Huang et al. [2]), which covers the primary minimum (see our Fig. 2) and a part of the maximum phases of the light curve. However, no improvement of our solution has been achieved, since the secondary minimum is lacking, and, apparently, the slight 'ellipticity' effect of the giant's shape on the light curve has been lost in their normalization procedure of the data at maximum light. Moreover, the broad-band TESS filter (λ = 600 - 1000 nm) might not allow a flawless fit being computed at a single wavelength (790 nm), because of the significant color changes of this binary in the primary minimum. Anyway, the parameters of our solution (see tables 1 and 2) are well inside the ranges of our TESS data fits.

For the dwarf component the BM3 fit gives a $T_{eff}$ of 9800, 9300 and 9000 K in B, V and Ic, respectively, in comparison with giant's $T_{eff}$ of 4950 K. The unusual passband dependence of its $T_{eff}$ is interpreted here by chemical peculiarities at the surface of this obviously slowly rotating star. This is confirmed by the intrinsic color index $(B–V)_0$ of 0.15 for the dwarf, which is significantly increased against a normally expected value of from –0.01 to +0.06 for the $T_{eff}$ range of 9800 to 9000 K, according to table 5 of Pecaut et al. [7]. HD 19805 and HD 23387 are (magnetic) peculiar A/B-star exemples showing clear similarities to our dwarf's properties. Therefore, the discolored dwarf component is adopted to have stellar model's unbiased $T_{eff}$ of 9600 K.

**Acknowledgements :**

This research has made use of the Simbad and VizieR databases operated at the **C**entre de **D**onnées astronomique, **S**trasbourg, France. This paper also includes data collected with the TESS mission, obtained from the MAST data archive at the Space Telescope Science Institute (STScI).

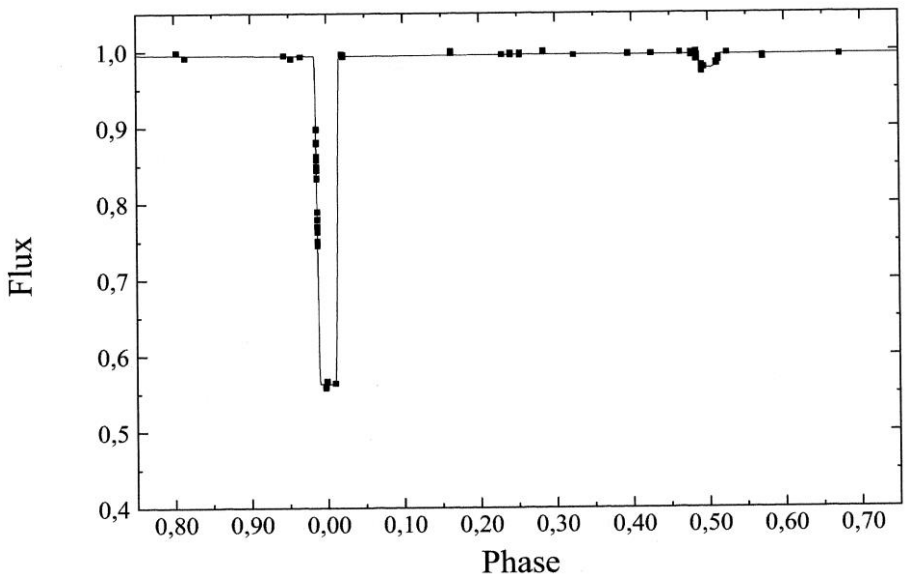

**Fig. 1 :** Computed light curve at λ$_{eff}$ 440 nm for our data in passband B

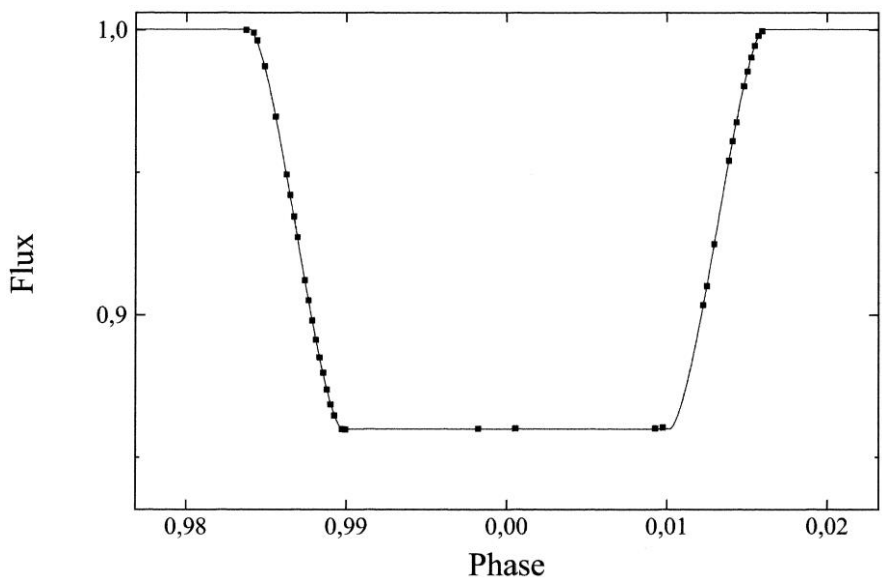

**Fig. 2 :** Computed light curve at λ$_{eff}$ 790 nm for TESS data of the primary minimum

**Table 1 : Parameters of binary system TYC 4481-358-1**

| | | |
|---|---|---|
| Epoch [BTJD] | 2458805.8596(4) | mid primary minimum TESS |
| Period [days] | 90.5288 | from VSX database |
| Maximum light in V [mag] | 9.9952(27) | from GAIA DR3 |
| Maximum light in B [mag] | 11.0998(48) | from GAIA DR3 |
| Minimum light in V [mag] | 10.322 | total eclipse (= giant's light) |
| Minimum light in B [mag] | 11.719 | total eclipse (= giant's light) |
| Secondary minimum in V [mag] | 10.020 | annular eclipse |
| Secondary minimum in B [mag] | 11.120 | annular eclipse |
| Eclipse duration [days] | 2.933 | |
| Orbital inclination i [deg] | 87.30 ± 0.10 | |
| Orbital radius a [$R_\odot$] | 148.7 ± 1.7 | for $R_\odot$ = 696342 km |
| Distance [pc] | 617.2 ± 5.6 | from GAIA DR3 |
| Extinction $A_V$ [mag] | 1.443 ± 0.024 | for excess $E_{B-V}$ = 0.403 |
| Total/selective extinction $R_V$ | 3.58 ± 0.06 | |

**Table 2 : Parameters of the components of TYC 4481-358-1**

| Parameter | giant | dwarf |
|---|---|---|
| Radius (R mean) [$R_\odot$] | 14.29 ± 0.16 | 2.39 ± 0.03 |
| Temperature $T_{eff}$ [K] | 4950 ± 110 | 9600 ± 200 |
| Ic-flux fraction at max. light | 0.8584 | 0.1416 |
| V-flux fraction at max. light | 0.7399 | 0.2601 |
| B-flux fraction at max. light | 0.5672 | 0.4328 |
| $(B–V)_0$ intrinsic color index | 0.993 | 0.150 |
| Mass [$M_\odot$] | 3.01 ± 0.10 | 2.39 ± 0.09 |

**References :**

[1] The International Variable Star Index (VSX) of the American Association of Variable Star Observers (AAVSO)
[2] C. X. Huang et al., 2020, Res. Notes AAS, 4, 204
[3] S. Wang & X. Chen, 2019, ApJ, 877, 116
[4] T. Lejeune et al., 1998, A&AS, 130, 65
[5] E.L. Fitzpatrick & D. Massa, 2009, ApJ, 699, 1209
[6] S. Ekström et.al., 2012, A&A, 537, A146
[7] M.J. Pecaut et al., 2013, ApJS, 208, 9